\documentclass{llncs}
\usepackage{amsmath,amssymb,epsfig,theorem}
\usepackage[ruled]{algorithm2e}
\usepackage{appendix}

\def\P{{\hbox{\bf P}}}
\def\E{{\hbox{\bf E}}}

 at 10 true pt

\def \small {\scriptsize}
\theoremstyle{plain}

\newcommand{\pc}[1]{}
\newcommand{\kc}[1]{}
\newcommand{\ptc}[1]{}

\newcounter{rnc}

\newenvironment{packed_item}{
\vspace{-0.5ex}
\begin{itemize}
  \setlength{\itemsep}{1pt}
}
{\end{itemize}\vspace{-1ex}}

{\begin{list}{}%
         {\setlength{\leftmargin}{#1}}%
         \item[]%
}
{\end{list}}

\title{Stochastic Matching with Commitment }
\author{Kevin P. Costello \inst{2} \and Prasad Tetali \inst{1} \and Pushkar Tripathi \inst{1} }

\institute{Georgia Institute of Technology\\
\and
 University of California at Riverside
}

\begin{document}
\maketitle
\allowdisplaybreaks
\thispagestyle{empty}
\begin{abstract}
We consider the following stochastic optimization problem first introduced by Chen et al. in \cite{chen}. We are given a vertex set of a random graph where each possible edge is present with probability $p_e$. We do not know which edges are actually present unless we scan/probe an edge. However whenever we probe an edge and find it to be present, we are constrained to picking the edge and both its end points are deleted from the graph. We wish to find the maximum matching in this model. We compare our results against the optimal omniscient algorithm that knows the edges of the graph and present a $0.573$ factor algorithm using a novel sampling technique. We also prove that no algorithm can attain a factor better than $0.898$ in this model.
\end{abstract}
\setcounter{page}{0}

\section{Introduction}
The matching problem has been a corner-stone of combinatorial optimization and has received considerable attention starting from the work of Jack Edmonds~\cite{edmonds}. There has been recent interest in studying stochastic versions of the problem due to its applications to online advertising and several barter exchange settings~\cite{Rapaport86,MM11,RRSJ97}. Much of the recent research focused on  studying matchings on {\em bipartite} graphs. In this paper we study a recently introduced variant on the stochastic online matching problem \cite{chen} on general graphs as described below. 

For {\bf $p$} a probability vector indexed by pairs of vertices from a vertex set $V$, let $G(V,p)$ denote  an undirected Erd\H{o}s-R\'{e}nyi graph on $V$.  That is, for any $(u,v) \in V \times V$, $p_{uv}=p_{vu}$ denotes the (known) probability that there is an edge connecting $u$ and $v$ in $G$. %, i.e. $p_{uv} = \Pr\left[ (u,v) \in E \right]$.
For every pair $(u,v) \in V \times V$ we are {\em not} told a priori whether there is an edge connecting these vertices, until we \textit{probe/scan} this pair. If we scan a pair of vertices and find that there is an edge connecting them we are constrained to \textit{pick} this edge and in this case both $u$ and $v$ are removed from the graph. However, if we find that $u$ and $v$ are not connected by an edge, they continue to be available (to others) in the future. The goal is to maximize the number of vertices that get matched.

We will refer to the above as the Stochastic Matching with Commitment Problem (SMCP), since whenever we probe a pair of adjacent vertices, we are committed to picking them.
\ptc{moved the next sentences to here; they used to be a bit later.}
%We compare our performance to the optimal off-line algorithm that knows the underlying graph and reports the maximum matching in it.
The performance of our algorithm is compared against the optimal offline algorithm that knows the underlying graph for each instantiation of the problem and finds the maximum matching in it. Note that since the input is itself random, the average performance of the optimal offline-algorithm is the expected size of the maximum matching in this random graph. We use the somewhat non-standard notation of $G(V,p)$, rather than $G_{n,p}$, since we will need to refer to the (fixed) vertex set $V$ and also since $p$ is a vector with typically different entries.
\ptc{threw in this clarification.}

\subsection{Our Results}
It is easy to see that the simple greedy algorithm, which probes pairs in an arbitrary order, would return a {\em maximal} matching in every instance of the problem and is therefore a factor $0.5$ approximation algorithm. We give a sampling based algorithm for this problem that does better than this:
\ptc{moved stuff around:}

\begin{theorem}
There exists a randomized algorithm that attains a competitive ratio of at least $0.573$ for the Stochastic Matching with Commitment Problem that runs in time $\tilde{O}(n^4)$ for a graph with $n$ vertices.  Furthermore, the running time can be reduced to $\tilde{O} (n^3)$ in the case where the expected size of the optimal matching is a positive fraction of the number of vertices in the graph.  
\end{theorem}
\kc{Maybe mention that the expected matching size can be estimated quickly?  I guess we need that in order to justify using the faster running time, since we need to know this when we choose our $C$} 

%
%The performance of our algorithm is compared against the optimal offline algorithm that knows the underlying graph for each instantiation of the problem and finds the maximum matching in it. Note that since the input is itself stochatic, given by an Erd\H{o}s-R\'{e}nyi graph $G(V,p)$, the average performance of the optimal offline-algorithm is the expected size of the maximum matching in this graph.
%
%
%The algorithm we present is non-adaptive  i.e. it determines an ordering on the edges then probes them according to this ordering ignoring edges one of whose end points is already matched. (KC: I don't believe this paragraph is accurate.  In particular, in the pruning stage we recompute the $q$ values in every step based on the edges we've seen so far, which seems pretty adaptive to me.)
%
%
%\pc{Commented some stuff according to KC. Refer to tex file for original stuff.}
%
%
Our algorithm uses \textit{offline simulations} to determine the relative importance of edges to decide the order in which to scan them. It is based on a novel sampling lemma that might be of independent interest in tackling online optimization settings, wherein an algorithm needs to make irrevocable online decisions with limited stochastic knowledge. This sampling trick is explained in section \ref{sec:sampling}. Even though the proof for our sampling lemma is based on solving an exponentially large linear program, we also give a fast combinatorial algorithm for it  (see Appendix \ref{app:combinatorial}).  As a result our algorithm can be implemented in time $O(n^3)$.

On the hardness front, we prove the following theorem, using  rigorous analysis of the performance of the {\em optimal} online algorithm for a carefully chosen graph.

%show an upper bound of $0.898$ for the performance of any algorithm for the SMCP.
\begin{theorem}
No algorithm can attain a competitive ratio better than $0.898$ for the SMCP.
\end{theorem}

\subsection{Previous Work}
The problem has similar flavor to several well known stochastic optimization problems such as the stochastic knapsack~\cite{DGJ05} and the shortest-path~\cite{BKMN06,PY91}; refer to~\cite{SS06} for a detailed discussion on these problems. SMCP also models several problems of practical significance which are discussed in Appendix \ref{app:applications}. We will now present more explicit connections between SMCP and several previously studied models of matching with limited information.

\textbf{Stochastic Matching Problems:} Chen et al.~\cite{chen} considered a more general model for stochastic matching than the one presented above. In their model every vertex $v \in V$ had a \textit{patience parameter} $t(v)$ indicating the maximum number of failed probes $v$ is willing to participate in. After $t(v)$ failed attempts, vertex $v$ would leave the system, and would not be considered for any further matches. Our model can be viewed as a special case of their setting where $t(v) = n$ for every vertex. However Chen et al., and subsequently Bansal et al.~\cite{bansal}, compared their performance against the optimal online algorithm. This was necessary because if we consider the case of the star graph, where each edge has a probability of $1/n$ and $t(v) = 1$ for every vertex $v$, then any online algorithm can match the center of the star with probability at most $1/n$, while there exists an edge incident on the center with probability $1-1/e$.  In contrast, our results are against the strongest adversary, i.e., the optimal offline omniscient algorithm. Clearly the performance of the optimal online algorithm can be no better than that of  such an  omniscient algorithm.
%, hence in this regard our results are a strict generalization of those presented in \cite{chen,bansal}.

In their model~\cite{chen}, Chen et al. presented a $1/4$ competitive algorithm,  which was later improved to a $0.5$ factor algorithm by Adamczyk~\cite{Adamczyk11}. Bansal et al.~\cite{bansal} studied the weighted version of the above model using a linear program to bound the performance of the optimal algorithm and gave a $1/4$ competitive algorithm for the general case, and a $1/3$ competitive algorithm for the special case of bipartite graphs. It is interesting to note that the linear program considered by Bansal et al. has an integrality gap of $2$ for general graphs which limits the best factor attainable by LP based algorithms. Another interesting aspect of their model is that the optimal online algorithm is a stochastic dynamic program having exponentially many states, and it is not even known if the problem is NP-hard when the patience parameters $t(.)$, can be arbitrary. \ptc{do we need to clarify: patience parameters?} \pc{italicized the initial definition, and added t(.) to clarify again}

\textbf{Online Bipartite Matching Problems:} Online bipartite matching was first introduced by Karp et al. in~\cite{KVV90}. Here one side of a bipartite graph is known in advance and the other side arrives online. For each arriving vertex we are revealed its neighbors in the given side. \pc{added prev line}The task is to match the maximum number arriving vertices. In~\cite{KVV90}, the authors gave a tight $1 - 1/e$ factor algorithm for this problem. This barrier of $1-1/e$ has been breached for various stochastic variants of this problem~\cite{FMMM09,BM08,MGS11}, by assuming prior stochastic knowledge about the arriving vertices.  Goel and Mehta \cite{GM08} studied the random order arrival model where the vertices of the streaming side are presented in a random order and showed that the greedy algorithm attains a factor of $1-1/e$. This was later improved to a $0.69$ competitive algorithm by \cite{MY11,KMT11}.

\begin{remark} The algorithm in~\cite{MY11,KMT11} can be thought of as the following randomized algorithm for finding a large matching in a given bipartite graph - randomly permute one of the sides and consider the vertices of the other side also in a (uniformly) random order. Match every vertex to the first available neighbor (according to the permutation) on the other side. It can be viewed as an oblivious algorithm that ignores the edge structure of the graph and can therefore be simulated in our setting. This yields a $0.69$ competitive algorithm  for the SMCP restricted to bipartite graphs.
\end{remark}

\textbf{Randomized Algorithms for Maximum Matchings:}
Fast randomized algorithms for finding maximum matchings have been studied particularly in the context of Erd\H{o}s-R\'{e}nyi graphs~\cite{AFP98,FP03,CFM10} starting from the work of Karp and Sipser~\cite{KarpSipser}. However all these algorithms explicitly exploit the edge structure of the graph and are not applicable in our setting. In \cite{ADFS95},  Aronson et al. analysed the following simple algorithm for finding a matching in a general graph - consider the vertices of the graph in a random order and match each vertex to a randomly chosen neighbor that is unmatched. This algorithm was shown to achieve a factor of $0.50000025$.

\begin{remark}
To the best of our knowledge, the algorithm in ~\cite{ADFS95} is the only non-trivial algorithm for finding a large matching in a general graph that works without looking at the edge structure. Since this algorithm works for \textit{arbitrary} graphs, it can be simulated in our setting and yields a $0.50000025$ factor algorithm for SMCP for general graphs. However we manage to improve the factor by exploiting the additional stochastic information available to us in our model.
\end{remark}

\subsection{Informal Description of the Proof Technique}
Observe that the simple algorithm, which weighs (or probes) an edge $e$ according to probability $p_e$, is not necessarily the best way to proceed. Consider a path having $3$ edges such that the middle edge is present with probability $1$ whereas the other two edges are each present with probability $0.9$.  Even though the middle edge is always present, it is unlikely to be involved in any maximum matching.  Conversely, the outer edges will always be a part of some maximum matching when they appear.

To determine the relative importance of edges, our algorithm relies on offline simulations. We sample from the given distribution to obtain a collection of representative graphs. We use maximum matchings from these graphs to estimate the probability (denoted by $q^*_e$) that a given candidate edge $e$ belongs to the maximum matching. Note that this is done as a preprocessing step {\em without} probing any of the edges in the given graph (a necessary requirement, as probing an edge could lead to unwanted commitments). Clearly the probability that a vertex would get matched in the optimal solution is the sum of $q^*_e$ for all edges incident on it and this gives us a way to approximate the optimal solution.

Similarly we can also calculate the conditional probability that an edge belongs to the maximum matching, given that it is present in the underlying graph. We use this as a measure of the importance of the edge. Observe that it is safe to probe edges where this conditional probability is large, since we are unlikely to make a mistake on such edges. After we are done probing these edges we are left with a residual graph where this conditional probability is small for every edge.

Ideally at this point what we would like to do is to simulate the fractional matching given by the $q^*_e$,  i.e., include every edge with probability $q^*_e$.  However, this is made impossible by the combination of our lack of knowledge of the graph and the commitments we are forced to make as we scan edges to obtain information about the graph.   To overcome these limitations, we devise a novel sampling technique, described in section~\ref{sec:sampling}, that gives us a partial simulation.  This sampling algorithm outputs a (randomized) ordering to scan the edges incident to a given vertex, so as to ensure that  edge $e$ is included with probability at least some large positive fraction of $q_{e}^*$. 

%Thus we can calculate, for every edge, the conditional probability that it belongs to the maximum matching given that it is present. Intuitively we use this as a measure of the importance of an edge. Clearly, it is safe to test edges where this conditional probability is large, since there is a small chance of making a mistake on such edges. This is achieved in the first stage of the algorithm, after which we are left with a residual graph where this conditional probability is small for all the remaining edges.

\section{Preliminaries}
\subsection{The Model}
We are given a set of vertices $V$, and for every unordered pair of vertices $u,v \in V$, we have a (known) probability $p_{uv}$  of the edge $(u,v)$ being present. These probabilities are independent over the edges. Let ${\mathcal D}$ denote this distribution over all graphs defined by $p$. Let $G(V,E)$ be a graph drawn from ${\mathcal D}$. We are given only the vertex set $V$ of $G$, but the edge-set $E$ is not revealed to us unless we \textit{scan} an unordered pair of vertices. A pair $(u,v) \in V \times V$ may be scanned to check if they are adjacent and if so then they are matched and removed from the graph. The objective is to maximize the expected number of vertices that get matched.

We compare our performance to the optimal off-line algorithm that knows the edges before hand, and reports the maximum matching in the underlying graph. We say an online algorithm ${\mathcal A}$ attains a competitive ratio of $\gamma$ for the SMCP if, for every problem instance ${\mathcal I} = (G(V,.),p)$, the expected size of the matching returned by ${\mathcal A}$ is at least $\gamma$ times the expected size of the optimal matching in the Erd\H{o}s R\'{e}nyi graph $G(V,p)$. That is, $\gamma = \displaystyle\min_{{\mathcal I} =  (G(V,.),p)} \left\{ \frac{E\left[ {\mathcal A}({\mathcal I}) \right]}{E\left[ \mbox{max matching in } G(V,p) \right]} \right\}$.

\kc{Definition collision here...we're using $\alpha$ for too many things.}\pc{ Fixed. Should we have diff variables in the definition of the corollary 1 and in the definition of algorithm(pruning stage) or can we reuse alpha here.}

\subsection{Definitions}
For any graph $H$ drawn from ${\mathcal D}$, let $M(H)$ be an arbitrarily chosen maximum matching on $H$.
%\kc{I changed this from ``lexicographically first``, because I'm not sure if the matching algorithms actually find the first matching or not.}\pc{Okay}
We define $$q^*_{uv}=\displaystyle\Pr_{H \leftarrow {\mathcal D}} \left( u \sim v \textrm{ in } M(H) \right).$$

Clearly $q^*_{uv} \leq p_{uv}$, since an edge cannot be part of a maximal matching unless it is actually in the graph.  In general, the ratio $q^*_{uv}/p_{uv}$ can be thought of as the conditional probability that an edge is in the matching, given that it appears in the graph.
For a given vertex $u$, the probability that $u$ is matched in $M$ is exactly
\begin{equation*}
 Q_u(G):=\sum_v q^*_{uv}.
\end{equation*}
This of course is at most $1$.  We will compare the performance of our algorithm against the expected size
of a maximum matching (denoted by OPT) for a graph drawn from ${\mathcal D}$. Thus we have,
\begin{equation}
 \label{eq:opt}
 \E[|{\rm OPT}|] = \E\left[ |M(H)|\right] = \frac{1}{2} \sum_u Q_u(G) = \sum_{(u,v)} q^*_{uv},
\end{equation}
where the last sum is taken over unordered pairs.  Finally define an unordered pair $(u,v)$ to be a \textit{candidate edge} if both $u$ and $v$ are still unmatched and ($u,v$) is yet to be scanned. At any stage let $F(G) \subseteq V \times V$ be the set of candidate edges, and for any $u \in V$, let $N(u,G) = \left\{ v \ | \ uv \in F(G) \right\}$. A vertex $u$ is defined to be {\em alive} if $|N(u,G)| > 0$.

\subsection{Sampling Technique} \label{sec:sampling}
\kc{Is there any way we can avoid getting into the technical stuff before actually stating the algorithm?}
\pc{I think the next 2 paragraphs that you added clarify the context. Should we add/remove more ?}
\ptc{I agree with KC, it might be less motivating to plunge into the sampling. Two options: (a) Either move the algorithm to here with a short preamble about $q_{u,v}$, or (b) insert a sentence here saying, the impatient/curious reader might read the description of the algorithm in Section~3 and return to the sampling part.}
\pc{added a line about the final outcome of this section}

In this section we will describe a sampling technique that will be an important component of our algorithm. A curious reader may directly read Corollary~\ref{cor:delta} and proceed to Section~\ref{sec:algo} to see an application of this technique.
Frequently over the course of our algorithm we will encounter the following framework: We have a vertex $v$, whose incident edges have known probabilities $p_{uv}$ of being connected to $v$.  We would like to choose an ordering on the incident edges to probe accordingly  so that each edge is included(scanned and found to be present) with some target probability of at least $r_{uv}$ (which may depend on $u$).

Clearly there are some restrictions on the $r_{uv}$ in order for this to be feasible; for example the situation is clearly hopeless if $r_{uv}>p_{uv}$.  More generally, for each subset $S$ of the neighborhood of $v$, it must be the case that the sum of the target probabilities of vertices in $S$ (the desired probability of choosing some member of $S$) is at most the probability that at least one vertex of $S$ is adjacent to $v$.  As it turns out, these are the \textit{only} necessary restrictions.

\begin{lemma}
\label{lem:non constructive}
 Let $A_1, A_2, \dots A_k$ be independent events having probabilities $p_1, \dots, p_k$.  Let $r_1, \dots, r_k$ be fixed non-negative constants such that for every $S \subseteq \{1, \dots ,k\}$ we have
\begin{equation} \label{eqn:samplingreq}
 \sum_{i \in S} r_i \leq 1-\prod_{i \in S} (1-p_i).
\end{equation}
Then there is a probability distribution over permutations $\pi$ of $\{1, 2, \dots, k\}$ such that for each $i$, we have
\begin{equation}
 \P(A_i \textrm{ is the earliest occurring event in } \pi) \geq r_i\,. \label{eq:sampling trick}
\end{equation}
\end{lemma}

\begin{proof}
 By the Theorem of the Alternative from Linear Duality~\cite{Farkas}, \ptc{a reference here, perhaps?} \pc{added}
 it suffices to show that the following system of $n!+1$ inequalities in $n+1$ variables $\{x_1, \dots, x_n, y\}$ does not have a non-negative solution:
\begin{eqnarray}
 y-\sum_k x_k r_k &<& 0 \label{eq:primal1}\\
 \forall \pi\in S_n, \ \ \ \ \ y-\sum_k x_k p_k \prod_{j<k \atop \textrm{in} \pi} (1-p_j) &\geq& 0 \label{eq:primal2}
\end{eqnarray}
Assume such a solution exists.  Without loss of generality we may assume $x_1 \geq x_2 \geq \dots \geq x_n \geq 0$.  Combining the first inequality with the inequality from the identity permutation, we have
\begin{equation} \label{eqn:dualeqn}
 \sum_{i=1}^n x_i p_i \prod_{j=1}^{i-1} (1-p_j)  < \sum_{k=1}^n x_k r_k.
\end{equation}

On the other hand, we have for each $k$ by applying \eqref{eqn:samplingreq} to $S=\{1, 2, \dots k\}$ that
\begin{equation*}
 \sum_{i=1}^k r_i \leq 1-\prod_{j=1}^k (1-p_j).
\end{equation*}
By weighting each of these equations by $(x_i-x_{i+1})$ and treating $x_{n+1}=0$ (note that each of these weights are nonnegative by assumption) and adding, we obtain
\begin{equation} \label{eqn:dualeqn2}
 \sum_{k=1}^n x_k r_k \leq \sum_{k=1}^n (x_k-x_{k+1})[1- \prod_{j=1}^k (1-p_j)].
\end{equation}
It can be checked directly that both the left side of \eqref{eqn:dualeqn} and the right hand side of \eqref{eqn:dualeqn2} are equal to

\begin{equation*}
 \sum_{S \subseteq \{1, 2, \dots n\} \atop S \neq \emptyset} (-1)^{|S|-1} x_{\max(S)} \prod_{i \in S} p_i,
\end{equation*}

 implying that the two equations contradict each other.  Therefore no such solution to the dual system can exist, so the original system must have been feasible.
\end{proof}

In theory, it is possible to find the desired distribution $\pi$ using linear programming.  However, it turns out there is a faster constructive combinatorial algorithm:

\begin{lemma}
\label{lem:constructive}
A probability distribution $\pi$ on permutations solving the program \eqref{eq:sampling trick} can be constructively found in time $O(n^2)$.
\end{lemma}
We defer the proof of this lemma and a description of the relevant algorithm to Appendix \ref{app:combinatorial}. The following corollary follows immediately from Lemma \ref{lem:constructive}. %\kc{General rule here...when referring to specific lemmas by number (Lemma 1, Lemma 2, etc.), capitalize Lemma} \pc{okay}

\begin{corollary}
\label{cor:delta}
Given a graph $G(V,E)$ and $u \in V$, such that $q^*_{uv}/p_{uv} < \alpha < 1$ for every $v \in N(u,G)$, there exists a randomized algorithm for scanning the edges in $\left\{ uv \ | \ v \in N(u,G)\right\}$ such any edge $uv$, $v \in N(u,G)$, is included in the matching with probability at least $\delta(u,G) q^*_{uv}$, where $$\delta(u,G) = \frac{1 -\exp(-\sum_{v \in N(u,G)} q^*_{uv}/\alpha)}{\sum_{v \in N(u,G)} q^*_{uv}}\,$$
\end{corollary}

\begin{proof}
Note that for any $u \in V$, and $S \subseteq N(u,G)$, $1-\prod_{v \in
S} (1-p_{uv}) \geq \sum_{v \in S} q^*_{uv}$, since the right side
represents the probability $u$ is matched to $S$ in our chosen maximal
matching and the left side the probability that there is at least one
edge connecting $u$ to $S$. Thus $(p,q^*)$ satisfy the condition for
Lemma \ref{lem:non constructive}. However, we can do better.  For any
given $S$, if we scale each $q_e$ by $\Bigl(1-\prod_{v \in S}
(1-p_{uv})\Bigr) / \sum_{v \in S} q^*_{uv}$, the above condition still
remains satisfied for that $S$.  Since $q^*_e/p_e < \alpha$ we have
\begin{subequations}
\begin{align}
\frac{1-\prod_{v \in S} (1-p_{uv})}{ \sum_{v \in S} q^*_{uv}} &\geq
\frac{1-\exp(-\sum_{v \in S} p_{uv})}{ \sum_{v \in S} q^*_{uv}}
%& \label{eq:delta1}
\geq \frac{1-\exp(-\sum_{v \in S} q^*_{uv}/\alpha )} {\sum_{v \in S}
q^*_{uv}} & \label{eq:delta2} \\
&\geq \frac{1-\exp(-\sum_{v \in N(u,G)} q^*_{uv}/\alpha)}{ \sum_{v \in
N(u,G)} q^*_{uv}} = \delta(u,G)\,,  \label{eq:delta3}
\end{align}
\end{subequations}
and \eqref{eq:delta3} follows since  $1-\exp(-\sum_{v \in S}
q^*_{uv}/\alpha ) / \sum_{v \in S} q^*_{uv}$ is a decreasing function
in $\sum_{v \in S} q^*_{uv}$,  thus achieving its minimum value at $S
= N(u,G)$. Therefore we can replace our $q^*$ by $\delta(u,G) q^*$ and
still have the conditions of Lemma \ref{lem:non constructive} hold.
\end{proof}

\section{Matching Algorithm on Unweighted Erd\H{o}s-R\'{e}nyi graphs}
\label{sec:algo}
Our algorithm can be divided into two stages.  The first stage involves several iterations each consisting of two steps - Estimation and Pruning. The parameters $\alpha$ and $C$ will be determined in Section \ref{sec:analysis}.
\begin{packed_item}
 \item \textbf{Step 1 (Estimation)}: Generate samples $H_1, H_2, \dots H_C$ of the Erd\H{o}s-R\'{e}nyi graph by sampling from ${\mathcal D}$.  For each sample, generate the corresponding maximum matching $M(H_j)$.  For every prospective edge $(u,v)$, let $q_{uv}$ be the proportion of samples in which the edge $(u,v)$ is contained in $M(H_j)$.  
\item \textbf{Step 2 (Pruning)}: Let $(u,v)$ be an edge having maximum (finite) ratio $q_{uv}/p_{uv}$.  If this ratio is less than $\alpha$, end Stage 1.  Otherwise, scan $(u,v)$.  If $(u,v)$ is present, add it to the partial matching; remove $u$ and $v$ from $V$, and return to Step 1; otherwise set $p_{uv}$ to $0$ and return to Step 1.  
\end{packed_item}
We recompute $q_{uv}$ every time we scan an edge. Stage $1$ ends when the maximum (finite) value of $q_e/p_e$ falls below $\alpha$. Note that at this point we stop recomputing $q$, and these values of $q$ will remain the same for each pair of vertices for the remainder of the algorithm. We now describe the second stage of the algorithm.

The second stage also has several iterations each consisting of two steps. At the start of this stage define $X = V$.
\begin{packed_item}
\item \textbf{Step 1 (Random Bipartition)}: Randomly partition $X$ into two equal sized sets $L$ and $R$ and let $B$ be the \textit{bipartite graph}  induced by $L$ and $R$.
\item \textbf{Step 2 (Sample and Match)}: Iterate through the vertices in $L$ in an arbitrary order, and for each vertex $u \in L$ sample a neighbor in $N(u,B)$ by choosing a vertex in $R$ using the sampling technique described in Corollary~\ref{cor:delta}\footnote{The algorithm described in Corollary~\ref{cor:delta} requires the exact estimates for $q^*_e$. However we will show in our analysis that for large enough samples $C$, $q_e$ defined above is a good estimate of $q^*_e$.}. At the end redefine $X$ to be the set of alive vertices in $R$ and discard the unmatched vertices in $L$. Recall that a vertex was defined to be alive if it is still unmatched and it has at least one candidate edge incident on it.
\end{packed_item}

The algorithm terminates when $X$ becomes empty.

%  The set of candidate edges that have one end point in $L$ and the other in $R$ induce a bipartite graph $B$, on $V$.
\kc{So are we recomputing the bipartition after every step?  Do we need to?} \pc{yes, since after each iteration in the second phase we would have no edges b/n L and R and now we wish to recurse over edges contained inside L and R.}
\ptc{So perhaps we clarify this by reviewing the definition of `alive' and mentioning the recursion?}\pc{added definition of 'alive' again}

\section{Analysis}
\label{sec:analysis}
\kc{ This whole section is treating both $q_{uv}$ and $\alpha$ like they're the actual probabilities of an event, but they're only approximations from our sample!  So, for example, OPT isn't really the sum of the q's, but it's close to it.  And the $\alpha$ in Lemma 3 is slightly different than the $\alpha$ we get from our sample.  I need to think a bit more about the best way to handle this in the exposition.  Maybe there should just be a separate section on approximations vs. actual?  This feels like it would lead to vague explanations as to how the argument should be adjusted though...}
\pc{I am not sure how much detail is needed for this...whether a vague statement about the errors being small/negligible is enough or we have to rigorously carryt the epsilons around in the whole proof.}
\kc{I added some stuff to take care of this, and it should be okay now modulo the comment at the end of the analysis about what to do when the expected matching size is $o(n)$.}

In this section we will analyze the competitive ratio for the algorithm described earlier. We begin by analyzing Stage $1$ of the algorithm. For each iteration in Stage $1$, define the residual graph at the start of the $i^{th}$ iteration to be $G_i$ starting with $G_1 = G$. We denote by $q_{e,i}^*$ the actual probability that $e$ is contained in the maximal matching on $G_i$ and $q_{e,i}$ as our estimate calculated in Step 1. We define
\begin{equation*}
\epsilon_e := \max_i |q_{e,i} - q_{e,i}^*|\,.
\end{equation*}
\kc{Is there a better notation here to make it clear  both $q_e$ and $q_e'$ change at each step?}
\pc{made it $q_{e,i}$.....this might be too much notation but it is only for one lemma, so it shuold be ok }
Let the total number of iterations in this stage be $k$ and let $G' = G_k$. Let $ALG_1$ be the set of edges that are matched in Stage $1$ and let $OPT(G_i)$ be the optimal solution in the residual graph at the start of the $i^{th}$ iteration. We have the following lemma.

\begin{lemma}
\label{lem:stage1}
$$\E\left[|OPT| - |OPT(G')|\right] \leq (2-\alpha) \E[|ALG_1|] + \sum_{e} \epsilon_e \,.$$
\end{lemma}

\begin{proof}
For $i \in [k]$, let $Gain(i)$ be $1$ if the edge scanned in the $i^{th}$ iteration is present, and $0$ otherwise. We will first show that
$E\left[|OPT(G_i)| - |OPT(G_{i+1})| \right] \leq (2 -\alpha) \E[Gain(i)]$. Three cases may arise during the $i^{th}$ iteration.
\begin{packed_item}
\item Case $1$: The edge scanned in the $i^{th}$ iteration is not present. Then $OPT(G_i) = OPT(G_{i+1})$ and $Gain(i) = 0$ thus, $|OPT(G_i)| - |OPT(G_{i+1})| = Gain(i) = 0$.
\item Case $2.1$: The edge scanned in the $i^{th}$ iteration is present but does not belong to $OPT(G_{i+1})$. This happens with probability $p_e-q_{e,i}^*$. Then $|OPT(G_i)| - |OPT(G_{i+1})| = 2$ and $Gain(i) = 1$.
\item Case $2.2$: The edge scanned in the $i^{th}$ iteration is present and belongs to $OPT(G_i)$. This happens with probability $q_{e,i}^*$. Then $|OPT(G_i)| - |OPT(G_{i+1})| = 1$ and $Gain(i) = 1$.
\end{packed_item}
Summing over all three cases, we see that
\begin{equation*}
\E [|OPT(G_i)| - |OPT(G_{i+1})|]= 2(p_e - q_{e,i}^*)+q_{e,i}^* \leq p_e (2-\alpha) + \epsilon_e\ , 
\end{equation*}
while the expected gain from scanning the edge is simply $p_e$.  The result follows from adding over all scanned edges, and noting for the additive factor that each edge is scanned at most once in the first stage (and indeed in the whole algorithm).
\end{proof}

\textbf{Analysis of Stage 2:} Let us begin by analyzing the first iteration of the second stage of the algorithm. The analysis for the subsequent iterations  would follow along similar lines. Let $G'$ be the residual graph at the start of the second stage, where $q_e/p_e < \alpha$ for every candidate edge $e$, and $1/2\sum_u Q_u(G') = \E[|OPT(G')|]$. The following lemma bounds the performance of the first iteration of Stage $2$ on $G'$. 

\begin{lemma}
\label{lem:first iteration}
The expected number of edges that are matched in the first iteration of Stage $2$ of the algorithm is at least $\left(1 - \frac{1}{e}\right) \left( 1 - e^{-1/2\alpha} \right) |OPT(G')| - \sum_e \epsilon_e $.
%The expected number of edges that are matched in the first iteration of stage $2$ of the algorithm is at least $\left(1 - \frac{1}{e}\right) \left( 1 - e^{-1/2\alpha} \right) |OPT(G')|$.
\end{lemma}
\begin{proof}
Let us define an indicator random variable $Y_u$ for every $u \in V$ that is $1$ if $u \in R$ and $0$ otherwise and $\Pr[Y_u = 1] = \Pr[Y_u = 0] = 1/2$. Thus $\E[Y_u] = 1/2$. We will use $u \sim v$ to denote that $u$ and $v$ get matched. The expected number of vertices in $R$ that will get matched in the first iteration is given by
\begin{subequations}
\begin{align}
&\E\left[ \mbox{matched vertices in }\ R\right]& \label{eq:stage2.0} \\
&= \E_Y \left[ \sum_u Y_u \left\{ 1 - \prod_{v \neq u}\left( 1 - Pr\left[ u \sim v , \ v \in L \ | \ u \in R\right] \right) \right\}\right]& \label{eq:stage2.1} \\
&= \E_Y \left[ \sum_u Y_u \left\{ 1 - \prod_{v \neq u}\left( 1 - Pr\left[ u \sim v  \ | \ u \in R, \ v \in L \right] \left(1-Y_v\right) \right) \right\}\right]& \label{eq:stage2.2} \\
&\geq \E_Y \left[ \sum_u Y_u \left\{ 1 - \prod_{v \neq u}\left( 1 - q_{uv}\delta(v,B \ | \ u \in R, v \in L)(1-Y_v) \right) \right\}\right]& \label{eq:stage2.3} \\
&\geq \E_Y \left[ \sum_u Y_u \left\{ 1 - \prod_{v \neq u}\left( 1 - q_{uv}(1-Y_v)\frac{ 1 - \left( 1-q_{uv}/\alpha \right)\displaystyle\prod_{w\neq u \neq v} \left( 1 - q_{vw}Y_w/\alpha \right)}{\sum q_{vw}Y_w + q_{uv}} \right) \right\}\right]& \label{eq:stage2.4}
\end{align}
\end{subequations}

\eqref{eq:stage2.2} follows from conditional probability, and \eqref{eq:stage2.3} and \eqref{eq:stage2.4} follow from Corollary \ref{cor:delta}, concerning our sampling technique. Since each of the random variables $Y_u$ are chosen independently, \eqref{eq:stage2.4} can be simplified as below.

\begin{subequations}
\begin{align}
&\E\left[ \mbox{matched vertices in }\ R\right]& \label{eq:stage2.01} \\
&\geq \sum_u  \E_Y\left[ Y_u \right] \E_Y \left[  1 - \prod_{v \neq u}\left( 1 - q_{uv}(1-Y_v)\frac{ 1 - \left( 1-q_{uv}/\alpha \right)\displaystyle\prod_{w\neq u \neq v} \left( 1 - q_{vw}Y_w/\alpha \right)}{\displaystyle\sum_{w\neq u \neq v} q_{vw}Y_w + q_{uv}} \right) \right]& \label{eq:stage2.5} \\
&= \frac{1}{2} \sum_u  \E_Y \left[  1 - \prod_{v \neq u}\left( 1 - q_{uv}(1-Y_v)\frac{ 1 - \left( 1-q_{uv}/\alpha \right)\displaystyle\prod_{w\neq u \neq v} \left( 1 - q_{vw}Y_w/\alpha \right)}{\displaystyle\sum_{w\neq u \neq v} q_{vw}Y_w + q_{uv}} \right) \right]& \label{eq:stage2.6}
\end{align}
\end{subequations}

In the following, \eqref{eq:stage2.7} can be derived from \eqref{eq:stage2.6} by using the identity $1-x < e^{-x}$, and  \eqref{eq:stage2.8} is obtained by noting that $(1-1/e)x < 1 - e^{-x}$ for $x \in [0,1]$.

\begin{subequations}
\begin{align}
&\E\left[ \mbox{matched vertices in }\ R\right]& \label{eq:stage2.02} \\
&\geq \frac{1}{2} \sum_u  \E_Y \left[  1 - \exp\left(  - \sum_{v \neq u} q_{uv}(1-Y_v)\frac{ 1 - \left( 1-q_{uv}/\alpha \right)\displaystyle\prod_{w\neq u \neq v} \left( 1 - q_{vw}Y_w/\alpha \right)}{\displaystyle\sum_{w\neq u \neq v} q_{vw}Y_w + q_{uv}} \right) \right]& \label{eq:stage2.7} \\
&\geq \frac{1}{2} \sum_u  \E_Y \left[ \left(1 - \frac{1}{e}\right)  \left(\sum_{v \neq u} q_{uv}(1-Y_v)\frac{ 1 - \left( 1-q_{uv}/\alpha \right)\displaystyle\prod_{w\neq u \neq v} \left( 1 - q_{vw}Y_w/\alpha \right)}{\displaystyle\sum_{w\neq u \neq v} q_{vw}Y_w + q_{uv}} \right) \right]& \label{eq:stage2.8} \\
&= \frac{1}{2} \left(1 - \frac{1}{e}\right) \sum_u  \E_Y \left[   \sum_{v \neq u} q_{uv}(1-Y_v)\frac{ 1 - \left( 1-q_{uv}/\alpha \right)\displaystyle\prod_{w\neq u \neq v} \left( 1 - q_{vw}Y_w/\alpha \right)}{\displaystyle\sum_{w\neq u \neq v} q_{vw}Y_w + q_{uv}} \right]& \label{eq:stage2.9}
\end{align}
\end{subequations}

Next observe that both $1-Y_v$ and $\frac{ 1 - \left( 1-q_{uv}/\alpha \right)\displaystyle\prod_{w\neq u \neq v} \left( 1 - q_{vw}Y_w/\alpha \right)}{\displaystyle\sum_{w\neq u \neq v} q_{vw}Y_w + q_{uv}}$ are decreasing convex functions in $Y$, thus their product is also a decreasing convex function.  Our next set of simplifications are as follows. Since for any multi-variate convex function $f$, $\E[f(y)] \geq f(\E[y])$ we can lower bound \eqref{eq:stage2.9} by \eqref{eq:stage2.10} below. Again \eqref{eq:stage2.11} is minimized when each of the $q_{vw}$'s are equal. Substituting this and simplifying we get \eqref{eq:stage2.13}. Finally $\frac{1-e^{-Q_v(G')/\alpha}}{Q_v(G')}$ is a decreasing function in $Q_v(G')$ that attains its minimum value when $Q_v(G') = 1$. Putting this value in \eqref{eq:stage2.14} gives \eqref{eq:stage2.15}. Further simplification yields the desired result.
\begin{subequations}
\begin{align}
&\E\left[ \mbox{matched vertices in }\ R\right]& \label{eq:stage2.03} \\
&\geq \frac{1}{2} \left(1 - \frac{1}{e}\right) \sum_u  \sum_{v \neq u} q_{uv}(1-\E[Y_v])\frac{ 1 - \left( 1-q_{uv}/\alpha \right)\displaystyle\prod_{w\neq u \neq v} \left( 1 - q_{vw}\E[Y_w]/\alpha \right)}{\displaystyle\sum_{w\neq u \neq v} q_{vw}\E[Y_w] + q_{uv}}& \label{eq:stage2.10}\\
&\geq \frac{1}{2} \left(1 - \frac{1}{e}\right) \sum_u  \sum_{v \neq u} \frac{q_{uv}}{2}\frac{ 1 - \left( 1-q_{uv}/\alpha \right)\displaystyle\prod_{w\neq u \neq v} \left( 1 - q_{vw}/2\alpha \right)}{\displaystyle\sum_{w\neq u \neq v} q_{vw}/2 + q_{uv}}& \label{eq:stage2.11}\\
&\approx \frac{1}{2} \left(1 - \frac{1}{e}\right) \sum_u  \sum_{v \neq u} \frac{q_{uv}}{2}\frac{ 1 - \displaystyle\prod_{w \neq v} \left( 1 - q_{vw}/2\alpha \right)}{\displaystyle\sum_{w \neq v} q_{vw}/2 }& \label{eq:stage2.12}\\
&\geq \frac{1}{2} \left(1 - \frac{1}{e}\right) \sum_u  \sum_{v \neq u} q_{uv}\frac{ 1 - \exp \left( -\displaystyle\sum_{w \neq v} q_{vw}/2\alpha \right)}{\displaystyle\sum_{w \neq v} q_{vw}/2 }& \label{eq:stage2.13}\\
&\geq \frac{1}{2} \left(1 - \frac{1}{e}\right) \sum_u  \sum_{v \neq u} q_{uv}\frac{ 1 - \exp \left( -Q_v(G')/2\alpha \right)}{Q_v(G')}& \label{eq:stage2.14}\\
&\geq \frac{1}{2} \left(1 - \frac{1}{e}\right) \sum_u  \sum_{v \neq u} q_{uv}\left( 1 - e^{ -1/2\alpha } \right) & \label{eq:stage2.15}\\
&\geq \frac{1}{2} \left(1 - \frac{1}{e}\right) \left( 1 - e^{-1/2\alpha} \right) \sum_u  \sum_{v \neq u} q_{uv}& \label{eq:stage2.16}\\
&\geq \frac{1}{2} \left(1 - \frac{1}{e}\right) \left( 1 - e^{-1/2\alpha} \right) \sum_u  Q_u(G') & \label{eq:stage2.17} \\
&\geq \left(1 - \frac{1}{e}\right) \left( 1 - e^{-1/2\alpha} \right) |OPT(G')| -\sum_e \epsilon_e\,. & \label{eq:stage2.18}
\end{align}
\end{subequations}
\end{proof}

For ease of notation, let $\phi = \left(1 - 1/e\right) \left( 1 - e^{-1/2\alpha} \right)$. Let $ALG_2$ be the set of edges that get matched in Stage 2 of the algorithm. The following lemma lower bounds $\E[|ALG_2|]$.
\begin{lemma}
\label{lem:alg2}
 $$\E[|ALG_2|] \geq \E[|OPT(G')|] \left[ \frac{\phi}{1- (\frac{1-\phi}{2})^2} \right] - \sum_e \epsilon_e\,.$$
\end{lemma}
\begin{proof}
Observe that not all candidate edges in $G'$ have been considered during the first iteration of Stage $2$. In particular, candidate edges with both end points in $R$ are yet to be considered. For analyzing the subsequent iterations in Stage $2$, we will  consider only these candidate edges. Clearly this only lower bounds the performance of the algorithm.

Analyzing \eqref{eq:stage2.17}, we notice that we have in fact proved something stronger in Lemma \ref{lem:first iteration}, i.e., we have shown that every vertex $v \in R$ is chosen with probability at least $\phi \ Q_v(G')$. By slightly altering the algorithm it is easy to ensure that for every $v\in R$, it is chosen with exactly this probability. Thus any vertex in $R$ survives the first iteration with probability $1 - \phi Q_v(G') > 1 - \phi$. Since the partitions $L$ and $R$ are chosen at random, the probability that a vertex is in $R$ and unmatched after the first iteration is at least $\mu = (1-\phi)/2$. Continuing this argument further, the probability that an ordered pair ($u,v$) is a candidate edge at the start of the $i^{th}$ iteration is the probability that both $u$ and $v$ have always been in $R$ in all previous iterations, and are still unmatched; this probability is at least $\mu^{2(i-1)}$.

Let $G'_i$ be the residual graph at the start of the $i^{th}$ iteration in Stage $2$, with $G'_1 = G'$. By the above observation and using linearity of expectation, the expected sum of $q_e$'s on candidate edges in $G'_i$ is lower bounded by $\mu^{2i-2} \sum_{e \in F(G')} q_e $.  Appealing to a similar analysis as in (the proof of) Lemma \ref{lem:first iteration}, the expected size of the matching returned by the $i^{th}$ iteration in the second stage is at least $\phi \mu^{2i-2} \sum_{e \in F(G')} q_e $. Summing over all iterations in Stage $2$ we have,

\begin{subequations}
\begin{align}
\E[|ALG_2|] &\geq \sum_i \phi \mu^{2i-2} \displaystyle\sum_{e \in F(G')} q_e 
%& \label{eq:alg2.1} \\
%&
\geq \phi \displaystyle\sum_{e \in F(G')} q_e  \sum_{i=1} \mu^{2i-2} &\label{eq:alg2.2} \\
&= \phi \displaystyle\sum_{e \in F(G')} q_e \frac{1}{1- \mu^2} \ \ 
%&\label{eq:alg2.3} \\
%&
=\ \  \frac{1}{2} \displaystyle\sum_{u \in G'} Q_u(G')  \frac{\phi}{1- \mu^2} &\label{eq:alg2.4} \\
&=  |OPT(G')| \frac{\phi}{1- \mu^2} \ \ 
%&\label{eq:alg2.5}  \\
%&
= \ \  |OPT(G')| \frac{\phi}{1- \left(\frac{1-\phi}{2}\right)^2}\,. & \label{eq:alg2.6}
\end{align}
\end{subequations}
\end{proof}

Now all that is left is to balance the factors for both the stages and set the optimal value of $\alpha$. In the subsequent theorem we find the optimal value of $\alpha$.
\begin{theorem} \label{thm:finalbound}
The above algorithm attains a factor of at least $0.573-2\gamma$ where $\sum_e \epsilon_e \leq \gamma \E(|OPT|)$.
\end{theorem}
\begin{proof}

By Lemma \ref{lem:stage1}, $\E[|OPT| - |OPT(G')|] \leq 2/(1+\alpha) \E[|ALG_1|]+\sum \epsilon_e$.
Also by Lemma \ref{lem:alg2}, $\E[|OPT(G')|] \leq \frac{1-(\frac{1-\phi}{2})^2}{\phi} \E[|ALG_2|]+\sum \epsilon_e$. Combining these two and substituting $\alpha = 0.255$ and $\phi = \left(1 - 1/e\right) \left( 1 - e^{-1/2\alpha} \right) = 0.543$ we have,
\begin{subequations}
\begin{align}
\E[|OPT|] &\leq (2 -\alpha) \E[|ALG_1|] + \frac{1-(\frac{1-\phi}{2})^2}{\phi} \E[|ALG_2|] +2 \sum_e \epsilon_e& \label{eq:thm1.1} \\
&= 1.74 (\E[|ALG_1|] + \E[|ALG_2|]) +2 \sum_e \epsilon_e& \label{eq:thm1.2} \\
&= 1.74 \E[|ALG|] + 2 \sum_e \epsilon_e& \label{eq:thm1.3}
\end{align}
\end{subequations}
Thus $\E[|ALG|]\ \geq 0.573\ \E[|OPT|]$.
\end{proof}
\kc{Fixed the rounding error, which slightly affects the bound, and added in the error term.  The $2$ in $0.573-2\gamma$ is superfluous (since no edge is scanned in both stages), but it might be more trouble than it's worth to explain this.  }

\subsection{Running Time Analysis}
In this section we will analyze the running time of our algorithm and determine the optimal value of parameter $C$. We will use $n$ to denote the number of vertices and $m$ to denote the number of edges that have a non-zero probability of being present.

By Lemma \ref{lem:constructive} it takes $O(n^2)$ time to probe the neighborhood of a given vertex, thus the second stage can be implemented in $O(n^3)$ time. Analysis of the first stage is slightly involved, since in this stage we wish to approximate $q^*_{e,i}$ by repeated sampling. The following lemma sets the optimal value of $C$ for which the total error caused by approximating $q^*_{e,i}$ by sampling is small. 

\begin{lemma}
\label{lem:C}
For $C=n \log^6(n)$ samples in Step 1, $\sum_e \epsilon_e$ is $o(n)$ with high probability.
\end{lemma}
\begin{proof}
We will give two separate bounds on the size of $\epsilon_e$. One will hold in the case where $q_{e,i}^*$ is not too small, the other for small $q_{e,i}^*$.  

\textbf{Bound 1:} For any given sample, we can think of the event $e \in M(H_j)$ as a Bernoulli trial with success probability $q^*_{e,i}$.  By Hoeffding's bound (\cite{Hoeff}, see Theorem 1.8 in \cite{TV06} for the specific formulation used), it follows that for any given edge and sample we have 
\begin{equation*} 
\P(|q_{e,i} - q_{e,i}^*| \geq \beta q_{e,i}^*) \leq \exp(-C q_{e,i}^* \zeta^2/4)\,.
\end{equation*}

This bound tells us that for any edge such that $C q_{e,i}^*$ tends to infinity sufficiently quickly, the maximum error coming from such an edge will likely be a tiny fraction of $q_{e,i}^*$.   However, it is possible that some $q_e$ could be exponentially small, so we cannot just take $C$ large enough so that all edges fall in this class.  We turn to the second bound for the remaining edges.

\textbf{Bound 2:} In this case we focus solely on the upper tail.  We know from the union bound that 
\begin{eqnarray*}
\P(q_{e,i} \geq q_{e,i}^*+\kappa) &\leq& \binom{C}{Cq^*_{e,i}+\kappa C} (q^*_{e,i})^{Cq^*_{e,i}+\kappa C} \\
&\leq& \left(\frac{eq_{e,i}^*}{q_{e,i}^*+\kappa}\right)^{Cq_{e,i}^*+\kappa C}\,,
\end{eqnarray*}
where the first inequality bounds the probability that the edge participates in at least $C(q_{e,i}^*+\kappa)$ matchings by the expected number of sets of $C(q_{e,i}^*+\kappa)$ matchings in which the edge participates.

Set $q_0 = \log^5 n/C$ where we will use bound $1$ for $q_{e,i}^* > q_0$ and bound $2$ otherwise. For $q^*_{e,i} > q_0$ using bound $1$, for an arbitrarily small constant $\zeta$ we have,
\begin{equation*}
\P(|q_{e,i} - q_{e,i}^*| \geq \zeta q_{e,i}^*) \leq \exp(-\log^5 n \zeta ^2/4) = o(1/n^4)\,.
\end{equation*}

Taking the union bound over all edges and trials and adding, we see
\begin{equation*}
\P(\sum_{q_{e,i}^* \geq q_0} \epsilon_e \leq \zeta \sum_e q_e^* )=1-o(1)\,.
\end{equation*}

Thus the total error accrued across all iterations is small with high probability.

Now let us set $\kappa=2 q_0$.  Applying the upper tail bound 2 above, we have for any $q_{e,i}^*<q_0$ that 
\begin{equation*}
\P(q_{e,i} \geq q_{e,i}^* + \kappa ) \leq \left( \frac{e q_{e,i}^*}{q_{e,i}^*+\kappa} \right)^{Cq_{e,i}^*+\kappa C} \leq (0.92)^{2 \log ^5 n}
\end{equation*}

The corresponding lower bound $q_{e,i} \geq q_{e,i}^*-\kappa$ follows trivially from the non-negativity of $q_e$.  

Taking the union bound over all samplings and all edges, we have that with high probability the total contribution to the error from this case is at most
$n^2 \kappa = n^2 \log^5 n / C$.  By Theorem \ref{thm:finalbound}, it suffices to make this a small fraction of the maximum expected matching.  Setting $C = n \log^6 n$ ensures that the total error is $o(n)$ (which suffices in the case where the matching is a constant fraction of all vertices, while setting $C = n^2 \log^6 n$ insures the error is $o(1)$ (which works in general) \footnote{If the expected maximum matching is itself $o(1)$ in size, then it follows from the independence of the edges that any edge which is scanned and found to be present is with high probability the \textit{only} edge in the graph!  So any algorithm trivially finds the maximal matching in such a graph.} 
\end{proof}

A naive implementation of the algorithm presented in section \ref{sec:algo} would require recalculating $q_e$ after every iteration in stage $1$. This can be quite time consuming since even the fastest implementation~\cite{MV80} of the maximum matching algorithm takes $O(m\sqrt{n})$ time. In the following lemma, we explain how to circumvent this bottleneck.  

\begin{lemma}
\label{lem:fast stage1}
Stage 1 of the algorithm can be implemented in $\tilde{O}(n^2 C )$ time.
\end{lemma}
\begin{proof}
Observe that we are not required to find the maximum matching in Step 1. We can instead work with a matching that is $1-\zeta$ ($\zeta$ is an arbitrary small constant) fraction of maximum matching and lose a small multiplicative factor in our analysis. This can be done in $O(m\log m)$ time using a result by Duan and Pettie et al.~\cite{DP10}. Also note that we can reuse the above matching across multiple iterations in Stage $1$. This is because the size of the maximum matching changes by at most $1$ across consecutive iterations. 

Concretely, we will modify the algorithm to calculate the approximate maximum matching using the algorithm in \cite{DP10} in $O(m\log m)$ time. Let $\Lambda$ be the estimate of the size of the maximum matching in the residual graph at any point during Stage $1$. We can probe up to $\Lambda\zeta$ edges that have $q_{e,i}/p_e > \alpha$ before recalculating $q_{e,i}$. This will induce only a small constant factor (function of $\zeta$) error in our analysis. Hence we would have at most $O(\log(m))$ iterations in Stage $1$ where we would be required to recompute $q_{e,i}$. Therefore we can implement Stage $1$ in $\tilde{O}(mC) = \tilde{O}(n^2C)$ time. 
\end{proof}

Finally, using Lemma \ref{lem:C} and \ref{lem:fast stage1} our algorithm can be implemented in $\tilde{O}(n^4)$ time.

On the hardness front, we prove the following theorem.
\begin{theorem}
\label{thm:upper bound}
No randomized algorithm can attain a competitive ratio better than $0.898$ for the SMCP.
\end{theorem}
\begin{proof}
Given any graph along with the probability $p_e$ for every edge the optimal algorithm can be found by writing a stochastic dynamic program. The states of the program are all possible subgraphs of the given graph and for each state we record the solution returned by the optimal algorithm. It is easy to see that such a dynamic program would have exponentially many states and would be quite infeasible to solve for the general problem. However it can be used to find the optimal algorithm for small examples.

We considered the complete graph on $4$ vertices where each edge is present with probability $p = 0.64$. In Lemma \ref{lem:opt algo k4} we evaluate the performance of the optimal online algorithm for this graph. Then in Lemma \ref{lem:exp max k4} we calculate the expected size of the maximum matching in this graph.

\begin{lemma}
The expected size of the matching found by the optimal online algorithm for SMCP for the Erd\H{o}s-R\'{e}nyi graph $G(4,0.64)$ is $1.607$.
\label{lem:opt algo k4}
\end{lemma}
\begin{proof}
Let us consider the complete graph $K_4$ as shown in Figure~\ref{fig:opt algo}(a). Without loss of generality we can assume that the optimal algorithm starts by scanning edge $ac$. If this edge is present, which happens with probability $p = 0.64$, then we are just left with one candidate edge. We can scan this edge next. Thus the contribution
to the expectation from this case is $p+p^2$.

Now suppose that $ac$ is not present, then we are left with the graph shown in Figure~\ref{fig:opt algo}(b). Clearly the optimal algorithm should not probe edge $bd$ since it can potentially lead to a smaller matching. Without loss of generality we may assume that the next edge to be scanned is $ab$. If this edge is present we are again left with one candidate edge $cd$ that is to be scanned. Hence in this case the contribution to the expectation is $(1-p)p(1 + p)$.

Otherwise we are left with the graph shown in Figure~\ref{fig:opt algo}(c). One can check that the optimal algorithm must next scan $bc$ or $ad$. Suppose it scans $bc$, and finds it to be present then we are again down to a single candidate $ad$ which can be scanned next. Therefore the contribution to the expected size of the matching is $(1-p)^2p(1+p)$.

If $bc$ is absent then the residual graph is shown in Figure~\ref{fig:opt algo}(d). Clearly for the graph in Figure~\ref{fig:opt algo}(d) the expected maximum matching is of size $1 - (1-p)^3$. The expected size of the matching returned in this case is therefore $(1-p)^3(1-(1-p)^3)$. Computing the expected value across all cases and substituting $p=0.64$ we find that the expected size of the matching returned by the optimal algorithm is $1.607$.

\begin{figure}[tbp]
	\centering
		\includegraphics[width=0.60\textwidth]{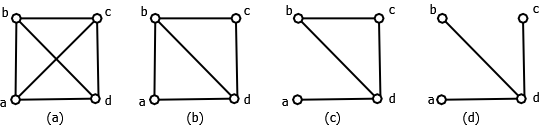}
		\caption{Intermediate Graphs}
	\label{fig:opt algo}
\end{figure}
\end{proof}

\begin{lemma}
The expected size of the maximum matching in the Erd\H{o}s-R\'{e}nyi graph $G(4,0.64)$ is $1.792$
\label{lem:exp max k4}
\end{lemma}
\begin{proof}
The given graph will not have any edge with probability $(1-p)^6$. It will have a matching of size $1$ for the graphs shown in figure \ref{fig:exp max} and their symmetric rotations.
\begin{figure}[htbp]
	\centering
		\includegraphics[width=0.60\textwidth]{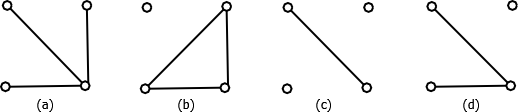}
		\caption{Graphs with unit size matching}
	\label{fig:exp max}
\end{figure}

This happens with probability $4p^3(1-p)^3 + 4p^3(1-p)^3 + 6p(1-p)^5 + 12p^2(1-p)^4$. In all other cases the graph will have a matching of size $2$ i.e. with probability $1 - (1-p)^6 - 8p^3(1-p)^3 - 6p(1-p)^5 - 12p^2(1-p)^4$. Thus the expected size of the maximum matching is $8p^3(1-p)^3 + 6p(1-p)^5 + 12p^2(1-p)^4 + 2(1 - (1-p)^6 - 8p^3(1-p)^3 - 6p(1-p)^5 - 12p^2(1-p)^4)$. Substituting $p = 0.64$ this is evaluates to $1.792$.
\end{proof}

Combining the results of Lemma \ref{lem:opt algo k4} and \ref{lem:exp max k4} we conclude that no online algorithm can achieve a factor
better than $1.607/1.792 = 0.898$ for the SMCP.
% is $1.792$. This establishes a lower bound of $1.792/1.607 = 1.115$ on the performance of any algorithm.
\end{proof}

\bibliographystyle{splncs_srt}
\bibliography{sampling}

%%%%%%%%%%%%%%%%%%%%%%%%%%%%%%%%%%%%%%%%%%%%%%%%%%%%%%%%%%%%%%%%%%%%%%%%%%%%%%%%%%%%%%%%%%%%%%%%%%%%%%%%%%%%%%%%%%%%%%%%%%%%%%%%%%%%%%%%%%%
%%%%%%%%%%%%%%%%%%%%%%%%%%%%%%%%%%%%%%%%%%%%%%%%%%%%%%%%%%%%%%%%%%%%%%%%%%%%%%%%%%%%%%%%%%%%%%%%%%%%%%%%%%%%%%%%%%%%%%%%%%%%%%%%%%%%%%%%%%%
%%%%%%%%%%%%%%%%%%%%%%%%%%%%%%%%%%%%%%%%%%%%%%%%%%%%%%%%%%%%%%%%%%%%%%%%%%%%%%%%%%%%%%%%%%%%%%%%%%%%%%%%%%%%%%%%%%%%%%%%%%%%%%%%%%%%%%%%%%%
%%%%%%%%%%%%%%%%%%%%%%%%%%%%%%%%%%%%%%%%%%%%%%%%%%%%%%%%%%%%%%%%%%%%%%%%%%%%%%%%%%%%%%%%%%%%%%%%%%%%%%%%%%%%%%%%%%%%%%%%%%%%%%%%%%%%%%%%%%%
%%%%%%%%%%%%%%%%%%%%%%%%%%%%%%%%%%%%%% BEGIN APPENDIX %%%%%%%%%%%%%%%%%%%%%%%%%%%%%%%%%%%%%%%%%%%%%%%%%%%%%%%%%%%%%%%%%%%%%%%%%%%%%%%%%%%%%
%%%%%%%%%%%%%%%%%%%%%%%%%%%%%%%%%%%%%%%%%%%%%%%%%%%%%%%%%%%%%%%%%%%%%%%%%%%%%%%%%%%%%%%%%%%%%%%%%%%%%%%%%%%%%%%%%%%%%%%%%%%%%%%%%%%%%%%%%%%
%%%%%%%%%%%%%%%%%%%%%%%%%%%%%%%%%%%%%%%%%%%%%%%%%%%%%%%%%%%%%%%%%%%%%%%%%%%%%%%%%%%%%%%%%%%%%%%%%%%%%%%%%%%%%%%%%%%%%%%%%%%%%%%%%%%%%%%%%%%
%%%%%%%%%%%%%%%%%%%%%%%%%%%%%%%%%%%%%%%%%%%%%%%%%%%%%%%%%%%%%%%%%%%%%%%%%%%%%%%%%%%%%%%%%%%%%%%%%%%%%%%%%%%%%%%%%%%%%%%%%%%%%%%%%%%%%%%%%%%

\begin{appendix}
\section{Applications of SMCP}
\label{app:applications}
Apart from its theoretical importance, SMCP models several important real-world scenarios. We briefly highlight some of these in this section.

\textbf{Kidney Exchange:} Consider a scenario where there are two incompatible donor/patient pairs where each donor is willing to donate a kidney for the patient, however is incompatible with her patient. In this setting we can perform a kidney exchange in which two incompatible
patient/donor pairs are identified such that each donor is compatible with the other pair's patient. Four simultaneous operations are then performed, exchanging the kidneys between the pairs in order to have two successful (compatible) transplants. Two donor/patient pairs can be tested to check if such an exchange is possible. Owing to the cost involved in testing and due to ethical concerns, it is desired that an exchange is performed whenever the test indicates that it is possible. We wish to maximize the number of such kidney exchanges.

This problem can easily be modeled as an instance of SMCP, where each donor and patient pair represents a node of the graph and the edges indicate possible pairs along  which this exchange is possible. Furthermore using external data, we can derive prior information about the likelihood of an exchange. We refer the reader to \cite{Rapaport86,RRSJ97} for more details.

\textbf{Online Advertising:} In display advertising, a user is shown ads during a browsing session. Whenever a user clicks on an ad, she is redirected to the merchant's website. The advertiser wishes to maximize the number of ads that get clicked across all users. This can be modeled as an instance of SMCP where the users and advertisers are the two sides of a bipartite graph. Whenever a user is shown an ad it is analogous to scanning the edge connecting them. Here the probability of an edge being present (an ad being clicked by a user) can be determined based on user profiles and is known to the advertiser in advance. \ptc{And ``Google" would like to maximize the matches? ;)}

\section{Combinatorial Algorithm for Implementing the Sampling Technique }
\label{app:combinatorial}
In this section we will give a complete proof of Lemma \ref{lem:constructive}.

By Lemma \ref{lem:non constructive}, we know that our constraint set is equivalent to
\begin{equation} \label{eqn:constraint}
\sum_{i \in S} r_i + \prod_{i \in S} (1-p_i) \leq 1
\end{equation}
holding for every $S$.  Suppose $S^*$ is a non-empty set for which the left hand side of \eqref{eqn:constraint} is maximized, and there is some $j_1 \in S$ and $j_2 \notin S$.  We would then have
\begin{eqnarray*}
\sum_{i \in S^*} r_i + \prod_{i \in S^*} (1-p_i) &\geq& \sum_{i \in S^* \backslash j_1} r_i + \prod_{i \in S^* \backslash j_1} (1-p_i) \\
\sum_{i \in S^*} r_i + \prod_{i \in S^*} (1-p_i) &\geq& \sum_{i \in S^* \cup \{j_2\}} r_i + \prod_{i \in S^* \cup \{j_2\}} (1-p_i)
\end{eqnarray*}
Rearranging both of the above inequalities yields
\begin{eqnarray*}
\prod_{i \in S^*} (1-p_i) &\leq& \frac{r_{j_1}}{\frac{1}{1-p_{j_1}} -1} \\
\prod_{i \in S*} (1-p_i) &\geq& \frac{r_{j_2}}{p_{j_2}}.
\end{eqnarray*}
Comparing these two inequalities, we have
\begin{eqnarray*}
\frac{r_{j_2}}{p_{j_2}} &\leq& \frac{r_{j_1}}{\frac{1}{1-p_{j_1}} -1} \\
&=& \frac{r_{j_1}(1-p_{j_1})}{p_{j_1}} \\
&\leq& \frac{r_{j_1}}{p_{j_1}}.
\end{eqnarray*}
Assume without loss of generality that the events are sorted in decreasing order by the ratio $r/p$.  By the above, we have proven
\begin{claim}
The left hand side of \eqref{eqn:constraint} is always maximized by $S=\{1, 2, \dots, k\}$ for some $k$.
\end{claim}
We also note the following:
\begin{claim}
If \eqref{eqn:constraint} is satisfied for all $S$ and tight for some $S_0$, then there is a solution to the program where any permutation having nonzero weight considers all variables in $S_0$ before any variable outside $S_0$.
\end{claim}
This is simply because the left hand side of \eqref{eqn:constraint} measures the sum of the probability that an event in $S_0$ is first and the probability that no event in $S_0$ occurs.  If the sum is $1$, then an event in $S_0$ must always be first if one is present, and that would not be possible if some other event could appear before it in $\pi$.

We now present our algorithm.

\textbf{Step 0:} (pre-processing:) Compute the largest $y$ for which the weights $(y r_i, p_i)$ still satisfy \eqref{eqn:constraint}.  If $y<1$, the constraints are infeasible.  If $y>1$, replace all $r_i$ by $yr_i$, at which point \eqref{eqn:constraint} is tight for at least one $S$.

This preprocessing step is only performed once.

\textbf{Step 1:} (slack removal) If there is no $k<n$ for which \eqref{eqn:constraint} is tight for $S=\{1, 2, \dots k\}$, compute the largest $z \leq 1$ such that the program remains feasible when for all $j$ we replace $r_j$ by
\begin{equation*}
r_j':=\frac{r_j - z p_j \prod_{i>j} (1-p_i)}{1-z}.
\end{equation*}
With probability $z$, consider the edges in decreasing order of index.  Otherwise, consider edges according to a distribution found by solving the problem with $r_j$ replaced by $r_j'$.  If $z=1$, we are finished.

\textbf{Step 2:} (divide-and-conquer) Find a $k<n$ for which \eqref{eqn:constraint} is tight for $S=\{1, 2, \dots, k\}$.  Solve, without pre-processing, the problems on $S$ (with the original target distribution) and $S^C$ (replacing the targets in $S^C$ by $r_j':=r_j / \prod_{i=1}^k (1-p_i)$).  Form the distribution by independently sampling from the distributions on $S$ and $S^C$ found by solving the subproblems, and consider all variables in $S$ before the variables in $S^C$.

For step $0$, note that multiplication by $y$ does not change the relative ordering of $r_i/p_i$.  This implies that the $y$ in question is the smallest $y$ for which one of the sets $\{1, 2, \dots, k\}$ makes \eqref{eqn:constraint} tight.  We can examine all such sets and find the $y$ in question in linear time by updating $\sum r_i$ and $\prod(1-p_i)$ each time $k$ increases to $k+1$.

For step $1$, note that
\begin{eqnarray*}
\frac{r_i'}{p_i}-\frac{r_{i+1}'}{p_{i+1}} &=& \frac{r_i - z p_i \prod_{j>i} (1-p_j)}{p_i} - \frac{r_i - z p_{i+1} \prod_{j>i+1} (1-p_j)}{p_{i+1}} \\
&=& \left(\frac{r_i}{p_i}-\frac{r_{i+1}}{p_{i+1}}\right) + z p_i \prod_{j>i+1} (1-p_j) \\
&\geq& 0.
\end{eqnarray*}
In other words, replacing $r_j$ by $r_j'$ again never alters the relative ordering of the ratios.  So again we only need consider $n$ sets to determine $z$, and can do this in linear time.

The $r_j'$ were chosen such that achieving a target of $r_j'$ with probability $(1-z)$ corresponds to achieving a target of $r_j$ in the original problem, so it is enough to solve this new problem.  The only claim that remains to be checked is that we can actually find the desired $k$ in Step $2$.  Since we know by our first claim that the left hand side of \eqref{eqn:constraint} is always maximized for some $S=\{1, 2, \dots, k\}$, it suffices to show that we can take $k<n$.

But for $S=\{1, \dots, n\}$, the left hand side of \eqref{eqn:constraint} is
\begin{eqnarray*}
\sum_{j=1}^n r_j' + \prod_{j=1}^n (1-p_j) &=& \frac{\sum_{j=1}^n r_j - z \sum_{j=1}^n p_j \prod_{i>j} (1-p_i)}{1-z} + \prod_{j=1}^n (1-p_j) \\
&=& \frac{\sum_{j=1}^n r_j - z \left(1-\prod_{j=1}^n  (1-p_j)\right)}{1-z} + \prod_{j=1}^n (1-p_j) \\
&=& \frac{\sum_{j=1}^n r_j + \prod_{j=1}^n (1-p_j) - z}{1-z} \\
&\leq& 1,
\end{eqnarray*}
where the last inequality comes from our assumption that \eqref{eqn:constraint} holds for $S=\{1,2,\dots,n\}$.  Intuitively, this corresponds to how imposing constraints on the order in which the events are placed (increasing $z$) has no effect on the probability at least one event occurs (the left hand side of \eqref{eqn:constraint} in the case where $S$ is everything).  So at the maximal $z$ (assuming $z<1$), some other constraint must also be tight, which gives us our $k$.

Since step $1$ takes at most linear time, it follows that the whole algorithm takes at most quadratic time.

\end{appendix}

\end{document}